\documentclass[12pt]{iopart}
\usepackage{graphicx}
\usepackage{color}
\usepackage{ulem}
\usepackage{iopams}
\expandafter\let\csname equation*\endcsname\relax
\expandafter\let\csname endequation*\endcsname\relax
\usepackage{cite}
\usepackage{amsmath}
\def\bea{\begin{eqnarray}}
\def\eea{\end{eqnarray}}
\def\ben{\begin{equation}}
\def\een{\end{equation}}
\def\benu{\begin{enumerate}}
\def\enu{\end{enumerate}}
\def\bei{\begin{itemize}}
\def\eei{\end{itemize}}

\def\sss{\scriptscriptstyle\rm}

\def\1var{(\bx_1...\bx\N)}

\def\br{{\bf r}}

\def\bx{{\br t}}

\def\bj{{\bf j}}

\def\s{_{\sss S}}

\def\N{_{\sss N}}

\def\sph_int{ {\int d^3 r}}

\begin{document}

\title{Ultrafast demagnetization in bulk vs thin films: an ab-initio study}
\author{ K. Krieger$^1$, P. Elliott$^1$, T. M{\"u}ller$^1$, N. Singh$^{1,2}$, J.K. Dewhurst$^1$, E. K. U. Gross$^1$ and S. Sharma$^{1,2}$}
\address{$^1$Max-Planck-Institut f\"ur Mikrostrukturphysik, Weinberg 2, D-06120 Halle, Germany.}
\address{$^2$Department of Physics, Indian Institute of Technology-Roorkee, 247667 Uttarakhand, India}

\date{\today}

\begin{abstract}
We report {\it ab-initio} simulations of the quantum dynamics of electronic charge and spins when subjected to intense laser pulses. By performing these purely electron-dynamics calculations for a thin film and for the bulk of Ni, we conclude that formation of surface has a dramatic influence of amplifying the laser induced demagnetization. The reason for this amplification is enhanced spin-currents on the surface of the thin films. We show that the underlying physics of demagnetization for bulk is dominated by spin-flips induced by spin-orbit coupling. In case of thin films the dominant cause of demagnetization is a combination of the flow of spin-currents and spin-flips. Furthermore, a comparison of our results with experimental data shows that below $\sim$120 fs processes of demagnetization is entirely dominated by purely electronic processes followed by which dissipative effects like Elliott-Yafet mechanism start to contribute significantly. 
\end{abstract}

\maketitle
\section{Introduction}
Femtomagnetism\cite{U09}, whereby the magnetic properties of a material are manipulated on the femtosecond timescale, was initiated by the experimental observation\cite{BMDB96,HMKB97,SBJE97,ABPW97} of ultrafast demagnetization of ferro-magnets subjected to an intense laser pulse. 
Due to the important technological implications of this phenomenon, e.g. in spintronics\cite{WABD01} or data storage\cite{TCKK04}, laser induced control of magnetism has since become a highly active field\cite{RVK00,GBB02,KKUP05,SKPM07,BVB09,BBHL10,KKR10,MRWP11,VGLS12,OBEC12,EBMP13,TLSG13,BLHH14,CMLC14,SVVK13,MMMF14,KMDS10,ZHLB09,CBOb11,BCO10}. 
As the devices utilizing ultra-short laser pulses to control magnetism hold the promise to reach the fastest possible electronic timescales, offering a speedup of several orders of magnitude over current state-of-the-art, magnetically operated devices, a large amount of work has gone into understanding the underlying physics of light-matter interactions. Among the most prominent suggested mechanisms are spin-orbit induced spin-dynamics\cite{KDES14,EKDS15,LH07,TP15}, all optical manipulation of spins\cite{VA09, MA14, EMD16}, Elliott-Yafet scattering\cite{KKKJ05,KMDS10,CBO11,ES11,CBLO13,IHF13}, Coulomb Exchange scattering\cite{ZHLB09} and Super-diffusive spin transport\cite{BCO10,BCO12}.

Most of the theoretical work to-date deals with the effect of laser pulses on the bulk of the magnetic material. Research that deals with interface/surface are model calculations\cite{CBOb11,BCO10} focusing only on a single aspect namely the diffusion of electrons across the interface. However, realistic devices contain surfaces, interfaces and bulk regions and a detailed understanding of the behaviour of these regions separately, under the influence of a laser pulse, is crucial\cite{ES16,JC16} to unraveling the complete physics of demagnetization. 
Experimentally it is a very challenging task to distinguish between contributions coming from various regions of the sample, but 
theoretically, by considering separate calculations for thin film and bulk of the same material, these effects can be disentangled. In the present work we use Ni thin film and bulk calculations to show that the underlying physics responsible for ultra-fast demagnetization in early femtoseconds is the same in both; spin-orbit induced spin-flips. Interestingly, we find that symmetry breaking originating from the formation of a surface (or interface) greatly enhances the demagnetization process. By explicitly treating spin and charge-currents in our simulations we find that the main reason behind this enhanced demagnetization in thin films is the presence of large spin-currents generated in these broken symmetry systems. 

In order to do this theoretical work we use the {\it ab-initio} method of time-dependent density functional theory (TDDFT)\cite{RG84,EFB09,TDDFTbook12,C11}, which is, in principle, an exact theory for studying light-matter interaction and makes no assumptions about the form of the electron dynamics.  
The study of dynamics of spins using TDDFT requires an extension where the magnetization density\cite{sdft} is treated as an unconstrained vector field. Such an extension was recently performed\cite{KDES14}, facilitating the present work. The calculations
presented here are purely electronic in nature and include contributions like (a) spin-
orbit induced spin-filps (b) restricted set of magnon excitations (by means of a super-cell calculations) and (c) spin-diffusion (or spin currents) to the process of demagnetization. Processes such as Elliott-Yafet scattering and electron-phonon (or lattice) induced spin-relaxation are ignored. A comparison with experiments then allows one to quantify the contribution of purely electronic processes to the physics of demagnetization and time scales at which other processes like Elliott-Yafet scattering become significant. In the present work we find that for times-scales below 120fs purely electronic processes dominate the physics of demagnetization. 

\section{Theoretical Aspects} 
Within TDDFT\cite{RG84} a Kohn-Sham(KS) Hamiltonian is used for time evolving the electronic wave-function. The scalar-relativistic KS Hamiltonian used in the present work reads:
\begin{eqnarray}
\label{hamil}
& &\hat{H}\s =\left[
\frac{1}{2}\left( \boldsymbol{\hat{p}} +\frac{1}{c}{\bf A}_{\rm ext}(t)\right)^2 +v_{s}({\bf r},t) \right.  \\ \nonumber
&+&\left. 
\frac{1}{2c} \boldsymbol{\hat{\sigma}} \cdot{\bf B}_{s}({\bf r},t) +
\frac{1}{4c^2} \boldsymbol{\hat{\sigma}} \cdot ({\boldsymbol \nabla}v_{s}({\bf r},t) \times \boldsymbol{[\hat{p}} + \frac{1}{c} \textbf{A}_{\rm ext} (t) ])\right],
\end{eqnarray}
where c is the speed of light, $\boldsymbol{ \hat{\sigma}}$ are the Pauli spin operators and ${\bf A}_{\rm ext}(t)$ is the vector potential representing an applied laser field. In the present work it is assumed that the wavelength of the applied laser pulse is much longer than the length of a unit cell. This assumption allows for the so-called dipole approximation and it implies that the spatial dependence of the vector potential ${\bf A}_{\rm ext}$ can be disregarded. The final term of Eq. (\ref{hamil}) is the spin-orbit coupling term which is included in the most general form and thus automatically includes any derived forms, e.g. Rashba-Dresselhaus coupling.
The KS effective potential $v_{s}({\bf r},t) = v_{\rm ext}({\bf r},t)+v_{\rm H}({\bf r},t)+v_{\rm xc}({\bf r},t)$ 
is decomposed into the external potential $v_{\rm ext}$, the classical electrostatic Hartree potential $v_{\rm H}$ and the 
exchange-correlation (XC) potential $v_{\rm xc}$. 
Similarly, the KS magnetic field is written as ${\bf B}_{s}({\bf r},t)={\bf B}_{\rm ext}(t)+{\bf B}_{\rm xc}({\bf r},t)$ 
where ${\bf B}_{\rm ext}(t)$ is the magnetic field of the applied laser pulse plus possibly an additional magnetic field 
and ${\bf B}_{\rm xc}({\bf r},t)$ is the XC magnetic field. In the present work we use the adiabatic local density 
approximation\cite{HK64,kubler} (ALSDA) for the XC functional.

In order to analyze the contribution of each term in this Hamiltonian (\ref{hamil}) to the dynamics of the magnetization ${\bf M}(t)= \int {\bf m}(\br,t) \mathrm{d}^3 r = \int \langle \boldsymbol{\hat{\sigma}} \hat{n}(\br,t)\rangle \mathrm{d}^3 r$, where $\hat{n}$ is the density operator and ${\bf m}(\br,t)$ is the magnetization density, we start by calculating the dynamics of the magnetization density using Ehrenfest's theorem:
\begin{equation}\label{e2}
\frac{\partial}{\partial t} m_j(\textbf{r}, t) 
= i \langle [\hat{H_s},\hat{\sigma}_j \hat{n}(\br,t)]\rangle,
\end{equation}
Substituting $\hat{H}_s$ from Eq. \ref{hamil} into Eq. \ref{e2} leads to
\begin{eqnarray} \nonumber
\label{e3}
\fl \frac{\partial}{\partial t} m_j(\textbf{r}, t) = 
 i \Big[ &\langle [ \frac{1}{2}(\hat{\boldsymbol{p}} + \frac{1}{c}\textbf{A}_{\rm ext}(t))^2, \hat{\sigma}_j \hat{n}(\br,t)] \rangle
 + \langle [\frac{1}{2c} \hat{\boldsymbol{\sigma}} \cdot \boldsymbol{\rm B}\s(\br,t),\hat{\sigma}_j \hat{n}(\br,t)] \rangle \\ \fl &+ \langle [\frac{1}{4c^2} \hat{\boldsymbol{\sigma}} \cdot (\boldsymbol{\nabla}v\s(\br,t) \times [\boldsymbol{\hat{p}} +\frac{1}{c} \textbf{A}_{\rm ext}]),\hat{\sigma}_j \hat{n}(\br,t)] \rangle \Big]. 
\end{eqnarray}
Evaluating the commutators results in
\begin{eqnarray} \nonumber
\label{dsdt}
\fl \frac{\partial}{\partial t} {\bf m}(\textbf{r}, t) &= 
- \boldsymbol{\nabla} \cdot \overleftrightarrow{\textbf{J}}(\br,t) + \frac{1}{c}[\textbf{B}_s (\br,t) \times \textbf{m} (\br,t)] + \frac{1}{4c^2} [\boldsymbol{\nabla} n(\br,t) \times \boldsymbol{\nabla} v\s(\br,t)]  \nonumber \\
\fl  & \hspace*{1cm}+ \frac{1}{2c^2} [\overleftrightarrow{\textbf{J}}^T(\br,t)-Tr\{\overleftrightarrow{\textbf{J}}(\br,t)\}] \cdot \boldsymbol{\nabla} v\s(\br,t),
\end{eqnarray}
where
$
\stackrel{\tiny{\mbox{$\leftrightarrow$}}}{\mathbf{J}}\!\!(\mathbf{r})
=\,\stackrel{\tiny{\mbox{$\leftrightarrow$}}}{\mathbf{J}}_p\!\!(\mathbf{r})\,
+\,\stackrel{\tiny{\mbox{$\leftrightarrow$}}}{\mathbf{J}}_d\!\!(\mathbf{r})
$
corresponds to the total spin-current tensor with paramagnetic component
$
\stackrel{\tiny{\mbox{$\leftrightarrow$}}}{\mathbf{J}}_p\!\!(\mathbf{r})
= \langle  \hat{\boldsymbol{\sigma}} \otimes \frac{1}{2} \{\hat{n}(\mathbf{r}), \hat{\mathbf{p}}\}\rangle
$
and diamagnetic component
$
\stackrel{\tiny{\mbox{$\leftrightarrow$}}}{\mathbf{J}}_d\!\!(\mathbf{r})
= \mathbf{m}(\mathbf{r}) \otimes \frac{1}{c} \mathbf{A}_{ext}
$. 
The change in the global moment, $\partial_t \textbf{M}(t)= \int \partial_t \textbf{m}(\br,t) \mathrm{d}^3 r$, can be evaluated by integrating Eq. \ref{dsdt}; the integral over $- \boldsymbol{\nabla} \cdot \overleftrightarrow{\textbf{J}}(\br,t)$ vanishes due to Gauss's law and the integral over $\frac{1}{4c^2} [\boldsymbol{\nabla} n(\br,t) \times \boldsymbol{\nabla} v\s(\br,t)]$ vanishes upon integrating by parts. 
For ALSDA $\textbf{B}_{\rm xc} (\br,t) \times \textbf{m} (\br,t)=0$  and in the absence of any external magnetic field (i.e. {\bf B}$_{\rm ext}$=0), the dynamics of the magnetization is given by:
\begin{eqnarray} \nonumber
\label{e5}
 \frac{\partial}{\partial t}\mathbf{M}(t)
  &=& \frac{1}{2c^2} \int d^3r \,
  [\stackrel{\tiny{\mbox{$\leftrightarrow$}}}{\mathbf{J}}\hspace{-0.2cm}\phantom{|}^T(\mathbf{r},t)
    - Tr\{\stackrel{\tiny{\mbox{$\leftrightarrow$}}}{\mathbf{J}}(\mathbf{r},t)\}] \cdot
  \boldsymbol{\nabla}v_s(\mathbf{r},t)
  \\[0.3cm]
  &=& \frac{1}{2c^2} \int d^3r
  \begin{bmatrix} \hat{x} \\ \hat{y} \\ \hat{z} \end{bmatrix} \boldsymbol{\times}
  \begin{bmatrix}
    \boldsymbol{\nabla}v_s(\mathbf{r},t) \times \mathbf{j}_x(\mathbf{r},t) \\
    \boldsymbol{\nabla}v_s(\mathbf{r},t) \times \mathbf{j}_y(\mathbf{r},t) \\
    \boldsymbol{\nabla}v_s(\mathbf{r},t) \times \mathbf{j}_z(\mathbf{r},t) \\
  \end{bmatrix}
\end{eqnarray}
with 
\begin{eqnarray}\nonumber
  \stackrel{\tiny{\mbox{$\leftrightarrow$}}}{\mathbf{J}}(\mathbf{r},t) =
  \begin{pmatrix} \mathbf{j}_x^T(\mathbf{r},t) \\ \mathbf{j}_y^T(\mathbf{r},t)
    \\ \mathbf{j}_z^T(\mathbf{r},t) \end{pmatrix}
\end{eqnarray} 
Here each spin-current density describes the flow of the respective spin-component. The $z$-component of Eq. \ref{e5} reads:
\begin{eqnarray}
\label{dmzdt}
  \frac{\partial}{\partial t}M_z(t) = \frac{1}{2c^2} \int d^3r \,
  [\boldsymbol{\nabla}v_s(\mathbf{r},t) \times \mathbf{j}_y(\mathbf{r},t)]_x -
  [\boldsymbol{\nabla}v_s(\mathbf{r},t) \times \mathbf{j}_x(\mathbf{r},t)]_y
\end{eqnarray}

\section{Results}
\begin{figure}[tbp]
\centerline{\includegraphics[width=\columnwidth,angle=-0]{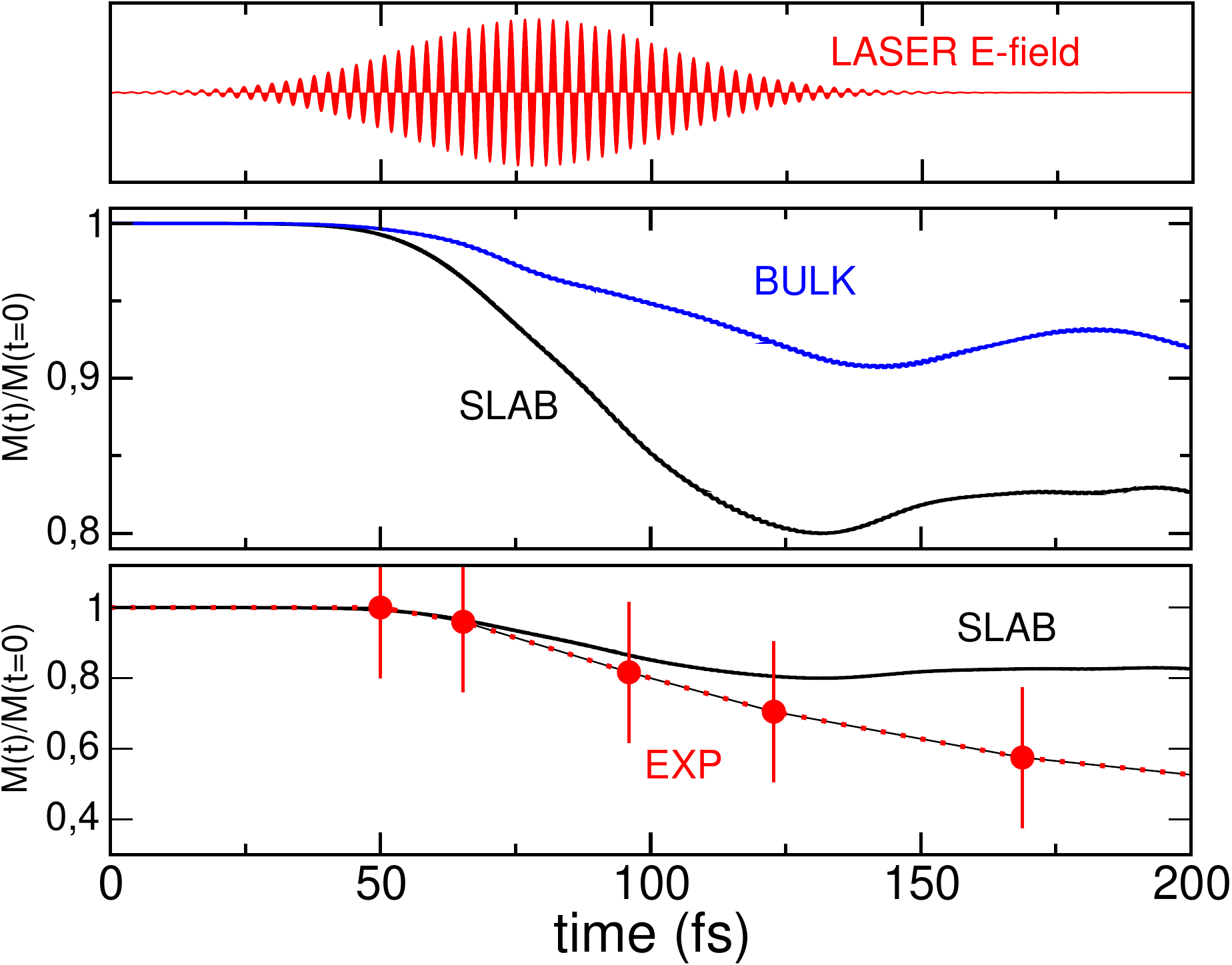}}
\caption{Top Panel: The electric field of the applied laser pulse with an intensity of $3.8\times 10^{11}$ W/cm$^2$, a fluence 8.05 mJ/cm$^2$ and a full width at half maximum (FWHM) of 40fs. Middle Panel: The dynamics of the $z$-component of the total magnetic moment for both the Ni thin film (black solid line) and bulk Ni (blue dashed line). Lower Panel: Comparison of the averaged layer-resolved moment to the experimental data of Ref. \cite{SKPM07}. The parameters of the pulse were also taken from this reference. }
\label{f:long}
\end{figure}
In order to distinguish the behaviour of spins on the surface (or interface) from those in the bulk of a sample subjected to an intense laser pulse, we study two cases in the present work; laser induced spin-dynamics in (a) bulk Ni and (b) a free standing film of Ni. Bulk Ni is easy to simulate in an electronic structure code which uses periodic boundary conditions (Elk code is used for all simulations\cite{elk}), but the computation of a thin film requires special care; in the present work we have used a $5$ atomic layer thick film with a 5 layer thick vacuum. The $z$-axis points in plane of the film and $y$-axis out of the plane. The ground state of this thin film is ferromagnetic with the magnetization pointing along the $z$-axis (in-plane of the film) (with a layer resolved moment of 0.71, 0.68 and 0.66 $\mu_{\rm B}$ starting from the layer adjacent to the vacuum). The pump laser pulse is then applied perpendicular onto this surface.  At $t=0$ we begin the simulation from the ground-state for both systems (bulk and film) and the results for the relative moment $M_z(t)/M_z(t=0)$ as a function of time are shown in  Fig. \ref{f:long}. 
In both cases, under the influence of the laser pulse of fluence 8.05 mJ/cm$^2$, the system is first optically excited (i.e. electrons are excited to energetically higher lying states), followed by a global loss in the magnetic moment. The two cases differ in the amount of demagnetization; the bulk shows a small loss of moment ($\sim 8\%$ during the simulation) while for the thin film this loss is much larger ($\sim 20\%$). Furthermore the calculations done in the film geometry can be compared to realistic experiments, where the laser pulse is applied perpendicular to the surface of the sample and such a comparison is made in Fig. \ref{f:long}c. The results show that we only reach an agreement with the XMCD data of Ref. \cite{SKPM07} for the first $\sim$120fs beyond which the theoretical work saturates and experimental data continues to demagnetize. 
The reason behind this disagreement between theory and experiment is the fact that in the present work we have included electronic only contribution (like spin-flips, spin-current and magnons) to the process of demagnetization and electron-lattice, electron-phonon, other Elliott-Yafet like mechanisms and radiation losses are ignored. These results thus point to two important findings –(a) how significant and at what time scales is the contribution of the electronic processes to the total (experimental) demagnetization and (b) the time scales at which dissipative processes like Elliott-Yafet mechanism start to be significant. We find that, in the present case, around $120$fs such processes start to contribute and become dominant for longer times.

At this point, it is natural to ask if the underlying mechanism for ultrafast demagnetization in bulk differs from a thin film. To answer this we note that in a purely electronic simulation there are two distinct spin excitation processes that can lead to a loss in the moment: processes that lead to local moment loss like magnons or non-collective canting of spins between atoms and processes that lead to global moment loss like spin-flips (or Stoner like excitations). 
We find that in both cases, for the first $\sim$120fs, the observed loss in magnetization along the $z$-direction is not accompanied by an increase in magnetization in the $x$ or $y$ directions, indicating that the long-range non-collinearity plays very little role on these time scales and the dynamics of Fig. \ref{f:long} is dominated by spin-flip processes for both bulk and thin films. Furthermore, the term in the Hamiltonian (Eq. \ref{hamil}) responsible for this magnetization dynamics is the spin-orbit coupling and setting this term to zero leads to no global demagnetization.

\begin{figure}[tbp]
\centerline{\includegraphics[width=\columnwidth,angle=-0]{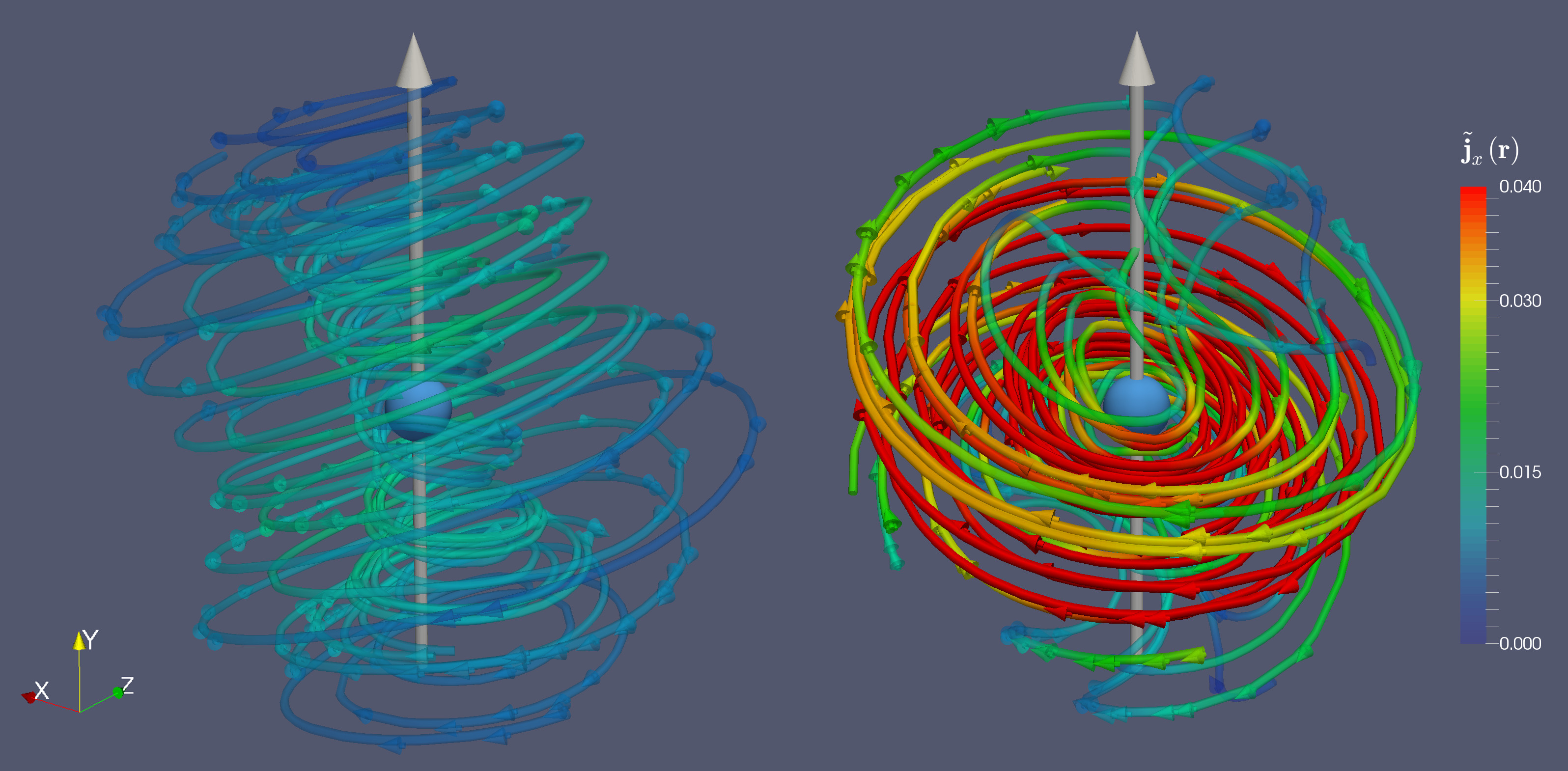}}
\caption{The vector field $\bj_x(t=120fs)$ around the Ni atom. The choice of $t=120$ fs is made based on the demagnetization being at its maximum. To easily visualize this vector field we show the streamlines in the left panel for bulk and in the right panel for Ni atom adjacent to the vacuum in the film.}
\label{f:Jx}
\end{figure}
Despite the similarity in the underlying physics, the amount of demagnetization is markedly different in the two cases (film and bulk). The question then arises: what leads to this strong difference?  From Eq. \ref{e5} it is clear that the spin-current tensor, $\stackrel{\tiny{\mbox{$\leftrightarrow$}}}{\mathbf{J}}$, is the sole quantity responsible for demagnetization. 
Hence to understand the difference between the thin film and the bulk demagnetization, in Fig. \ref{f:Jx}, we plot the $x$-spin-component of the spin-current, $\bj_x $, which can be defined following Eq. \ref{dsdt} as $\bj_x(\br)=\langle  \hat{\sigma}_x \otimes \frac{1}{2} \{\hat{n}(\mathbf{r}), \hat{\mathbf{p}} + \frac{1}{c} \mathbf{A}_{ext}\}\rangle = \langle  \hat{\sigma}_x \otimes \hat{\bj}\rangle $. This quantity can be interpreted as the vector field showing the flow of spins orientated along the $x$-direction. As an aid to visualizing these currents, we can create streamlines by following the direction of the vector at each point with the strength of this flow shown in colour.  Fig. \ref{f:Jx} thus shows how the spins pointing in the $x$-direction are flowing (circulating) around the $y$-axis. From these results it is clear that for both the bulk and the film the spin-current loops clockwise about the $y$-axis, as is required by Eq. \ref{dmzdt} to cause a change in the moment. However, the magnitude of the spin-current in the case of the thin film is much larger than for the bulk and this enhanced spin-current then leads to larger demagnetization in the film. The reason behind this we find to be the broken symmetry due to surface/interface formation, which allows for large surface currents due to the presence of localized electronic wave-function (a natural consequence of the lower symmetry). The flow of this surface current towards the center of the film then causes large spin-currents every where in the film resulting in a significant demagnetization.

\begin{figure}[t]
\centerline{\includegraphics[width=\columnwidth,angle=-0]{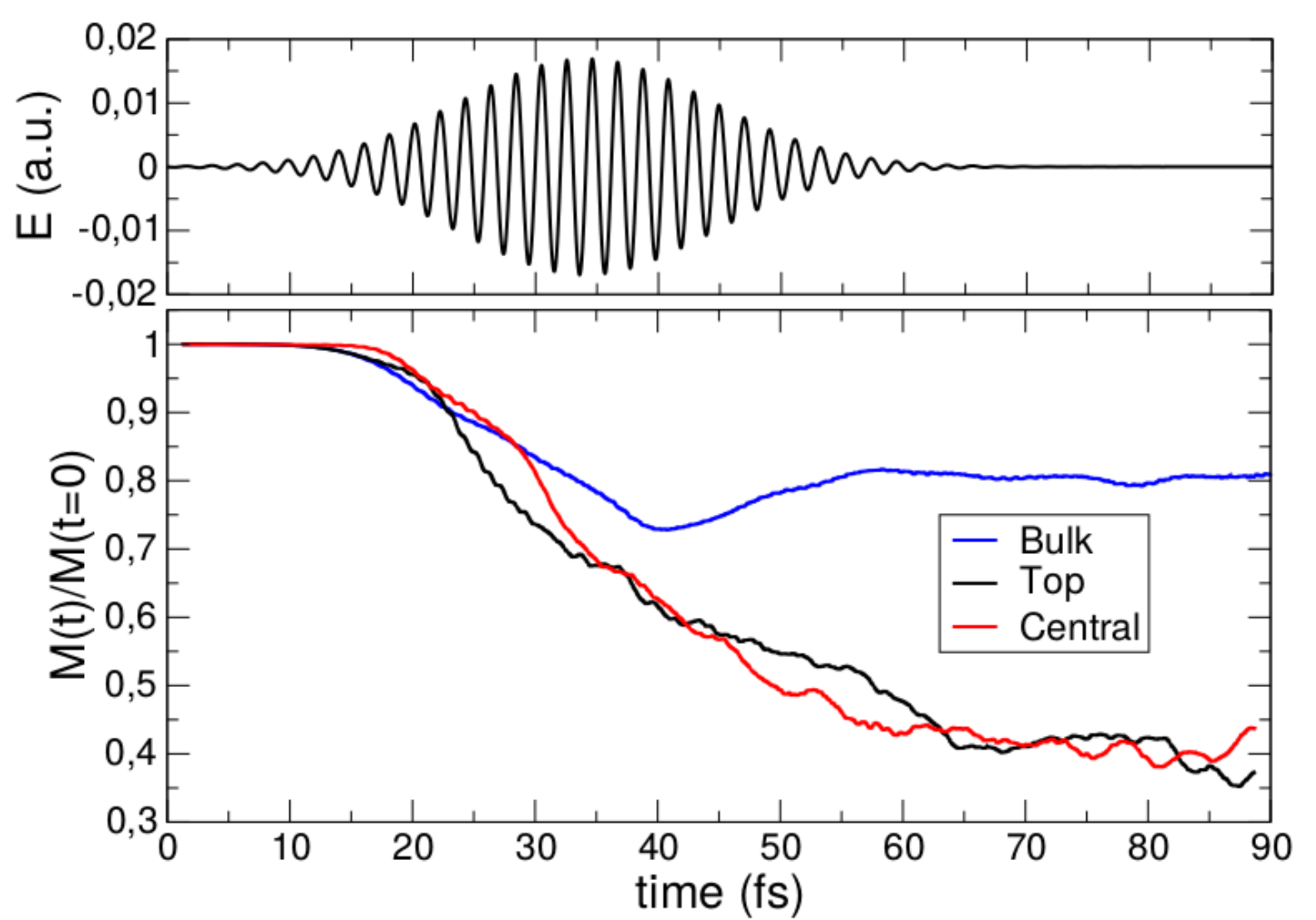}}
\caption{Top Panel: The electric field of the applied laser pulse, with peak intensity $1\times 10^{13}$ W/cm$^2$, FWHM of $17$ fs, and fluence of $91.27$ mJ/cm$^2$. 
Lower panel: The dynamics of the $z$-component of the magnetic moment for top (black) and central (red) layers of Ni film and for bulk Ni (blue). 
The shorter pulse leads to a significant increase of the observed demagnetization. Additionally, the demagnetization occurs much faster.}
\label{f:mid}
\end{figure}
To gather further insight into different behaviour of thin films and bulk it is instructive to compare the layer resolved magnetization of the film to that of the bulk. These results, for an ultra-short pulse with FWHM=$17$fs, are plotted in Fig. \ref{f:mid} for the bulk, the top layer of the film and the very central layer of the film. From this data it is clear that during the first $\sim$30 fs the demagnetization in the bulk and the central layer of the film are similar while the top layer of the film demagnetizes faster. Beyond $40$fs the layers of the film continue to demagnetize while the bulk saturates. 
These results raise two interesting questions;
why the demagnetization is almost the same for the bulk and the film in the first 30fs and why do the layers of the film continue to demagnetize whereas the bulk saturates beyond 40fs?
One of the major reasons for these effects is that below $\sim$30fs the spin-currents have almost the same magnitude for the bulk and the film, subsequently the broken symmetry allows for stronger currents in the film, while for the case of the bulk these currents stay small. In case of the film these spin-currents flow from one layer to another and continue to flow beyond the first  40fs, leading to further demagnetization of the layers of the film. This explains the observed physics of demagnetization in the two cases.  
It would be interesting to know the thickness at which the very central layer of the film start to behave like the bulk of Ni. But TDDFT, a state-of-the-art method to deal with light-matter interaction, comes at a very high computational cost. As a result, treating a film greater than a few atomic-layer thickness is not feasible with the resources available to us.

Finally, it is crucial to mention that in Fig. \ref{f:mid} we have used the predictive power of TDDFT to study the influence of ultra-short laser pulses on the spin dynamics. Presently there are no experiments reported which utilize such pulses, however the parameters are within the range of available technology.  The effect of this ultra-short pulse is faster and much enhanced demagnetization;  for a film $51\%$ of the moment is lost while for bulk this value is $17\%$. The corresponding values for the long pulse (see Fig. \ref{f:long}) are 20\% for the film and 8\% for the bulk.

\section{Discussion}
By performing {\it ab-initio} simulations for laser pulse excited ferromagnetic thin films, we have demonstrated that thin films show enhanced spin-orbit mediated demagnetization compared to a bulk. We have further demonstrated that this enhancement in the demagnetization near the surface of a film is due to the broken symmetry which increases the rotating spin currents in the system. These calculations show the importance of treating the spin-orbit coupling as well as charge- and spin-currents at the same footing.
   
From our calculations one can conclude that for a typical laser pump pulse (as currently used in  experiments), demagnetization (caused by purely electronic processes) is strong within the first few atomic layers of a material but will then decrease to the smaller bulk value as we get deeper into the sample and it becomes more bulk-like. For longer duration pulses, such as that used in Fig. \ref{f:long}, for the first $\sim$120fs purely electronic processes dominate the physics of demagnetization.
Beyond first 120fs, these electronic mechanisms will exist in addition to dissipative mechanisms such as Elliott-Yafet, and it will require more ingenious experiments to disentangle and distinguish them. This early time purely electronic regime is of great importance for optimal control and device production at ultrashort timescales as it allows for the possibility of coherent control of the electrons.

\section{Computational Details}
Considering future coherent control of spins by light, in the present work we explicitly concentrate on the electronic degrees of freedom during the early femtoseconds. Dissipative processes that induce decoherence such as radiation and phonons are not included (nuclei are kept fixed during the simulation). 
At this point it is important to mention that collective magnetic excitations (e.g. magnons) are also purely electronic in nature and have been included in the present work through usage of a large super-cell. However, we found that magnons only have a minor contribution to the spin-dynamics in the time scales studied in the present work\cite{KDES14}.

State-of-the art full potential linearized augmented plane wave (LAPW) method implemented within the Elk
code\cite{elk} is used in the present work. The core electrons (with Eigenvalues below 95eV) are treated using the
radial Dirac equation while higher lying electrons are treated using the scalar relativistic Hamiltonian in the presence 
of the spin-orbit coupling. To obtain the 2-component Pauli spinor states, the Hamiltonian containing only the scalar 
potential is diagonalized in the LAPW basis: this is the first-variational step. The scalar states thus obtained are then 
used as a basis to set up a second-variational Hamiltonian with spinor degrees of freedom\cite{singh}. This is more efficient
than simply using spinor LAPW functions, but care must be taken to ensure that there is a sufficient number
of first-variational eigenstates for convergence of the second-variational problem. In the present work
300 empty states per {\bf k}-point are used.

For Ni a face centered cubic crystal structure with lattice spacing of $3.52$ \AA  \, is used.  A {\bf k}-point grid of $8\times 8\times 8$ is used for the bulk and $1\times 8\times 8$ for the film calculations. A full geometry optimization for the Ni-film was performed, however, we found that for intense laser pulses, such as used in present work, geometry optimization does not change results significantly. The maximum augmented plane-wave cutoff 
for the orbitals was $3.5$ au$^{-1}$ corresponding to an energy cutoff of approximately $160$ eV, and for the density and potential a value of 12 au$^{-1}$ was used. An angular momentum cutoff of 8 for orbitals and 7 for densities and potential was used.
A time step of $0.002$ fs was used for time propagation\cite{algo}. The initial state for all TDDFT calculations was the DFT ground-state, which is at a temperature of 0K. We use the adiabatic local density approximation\cite{HK64,kubler} for the XC functional. The film calculations required $900000$ CPU hours, running on the HYDRA supercomputer at RZG Garching for $1$ month. 

\section{Acknowledgements}
TM and SS would like to thank QUTIF-SPP for funding. PE and SS would like to acknowledge funding from DFG through SFB762 project.


\end{document}